\documentclass[12pt]{JHEP3}
\usepackage{multicol}
\usepackage[dvips]{epsfig,graphics}
\usepackage{graphicx}
\usepackage{cite}

\newlength{\wth}
\newcommand{\ntlinoOne}{{{{\tilde{\chi}}_1^0}}}
\newcommand{\ntlinoTwo}{{{{\tilde{\chi}}_2^0}}}
\newcommand{\squark}{{{{\tilde{q}}}}}

\title{Three body kinematic endpoints in SUSY models with non-universal Higgs masses }
\author{ C. G. Lester, M. A. Parker and M. J. White\\ 
Cavendish Laboratory. Madingley Road. Cambridge CB3 0HE, UK
}
\abstract{We derive and present expressions for the kinematic endpoints that arise in the invariant mass distributions of visible decay products of cascade decays featuring a two body decay followed by a three body decay. This is an extension of a current technique that addresses chains of successive two body decays. We then apply these to a supergravity model with Non-Universal Higgs Masses (NUHM), having simulated a data set using the \tt ATLFAST \rm detector simulation. We find that, should such a model be chosen by nature, the endpoints will be visible in ATLAS data, and we discuss the problems associated with mass reconstruction in models with a similar phenomenology.}
\keywords{Beyond the Standard Model, mSUGRA, NUHM, Kinematic Endpoints, Mass Reconstruction}

\preprint{Cavendish-HEP-2005-21\\
          SN-ATLAS-2006-058}
\begin{document}
\section{Introduction}\label{intro}
The Standard Model (SM) of particle physics has been enormously successful in predicting a wide range of phenomena with great accuracy and precision. In spite of this, however, there are specific issues that lead one to conclude that the SM is only an effective theory, which is thus incapable of describing physics at arbitrarily high energies. Searches for new physics in the ATLAS detector of the Large Hadron Collider (LHC) will be as exciting as they are challenging.

There is a plethora of models on the market that attempt to solve the
problems of the SM, and supersymmetry is a relatively strong contender
(see \cite{Martin:1997ns} for a concise introduction). By generalising
the space-time symmetries of the gauge field theories of the SM to
include a transformation between fermion and boson states, one
introduces `superpartners' for the SM particles which differ from
their SM counterparts only by their spin if supersymmetry is unbroken;
in the broken case they differ also in mass. In the process, one
obtains a solution to the gauge hierarchy problem of the SM whilst
also ensuring gauge unification at high energy scales, provided that
the masses of the superpartners are below the TeV range. This provides
the major motivation for supersymmetry searches at the LHC, and much
effort has already been invested in developing strategies to measure
particle masses and SUSY parameters at ATLAS.

It is noted that an interesting feature of most supersymmetric models
is the existence of a multiplicatively conserved quantum number called
R-parity, in which each superpartner is assigned $R=-1$, and each SM
particle is assigned $R=+1$. R-parity conservation both ensures that
sparticles must be produced in pairs and forces the lightest
supersymmetric particle (LSP) to be absolutely stable. Thus, we obtain
a ideal dark matter candidate, and can use recent measurements of dark
matter relic density to impose constraints on SUSY models.

The lack of any current collider observations of sparticles means that
SUSY must be a broken symmetry, and our ignorance of the breaking
mechanism unfortunately means that the parameter set of the full
Minimal Supersymmetric Standard Model (MSSM) numbers 124 (where SUSY
breaking has given us 105 parameters on top of the SM).\footnote{Note
that the number of MSSM parameters mentioned in this section is true
only for the `old' (i.e.\ {\em pre}-neutrino oscillation) Standard
Model.}  The difficulty of exploring such a large parameter space has
meant that practically all phenomenological studies to date have been
performed in simplified models in which various assumptions are made
to reduce the parameter set to something of the order of 5, of which
two are fixed. A popular example is the minimal Supergravity breaking
scenario (mSUGRA), in which one unifies various GUT scale parameters,
obtaining the following parameter set: the scalar mass $m_0$, the
gaugino mass $m_{1/2}$, the trilinear coupling $A_0$, the ratio of
Higgs expectation values tan$\beta$, and the sign of the SUSY Higgs
mass parameter $\mu$.

Many ATLAS studies have been performed in the mSUGRA framework, using
both full and fast simulation
\cite{Allanach:2000kt,AtlasTDR,Lester:2001zx,Borjanovic,Lester:2005je,Ozturk,Lari,Comune},
and one obtains an interesting range of phenomenologies over the
parameter set. However, although one can devise relatively strong
theoretical motivations for gaugino mass universality at the GUT
scale, there is no reason why the soft supersymmetry-breaking scalar
masses of the electroweak Higgs multiplets should be universal, and it
is thus particularly important to consider models in which one breaks
the degeneracy of these masses. These Non-Universal Higgs Mass (NUHM)
models lead to yet more interesting phenomenological effects, and a
set of benchmark points consistent with current measurements of the
dark matter relic density, the $b \rightarrow s\gamma$ decay branching
ratio and $g_{\mu} -2 $ was presented in
\cite{DeRoeck:2005bw}. Furthermore, it was observed in
\cite{Baer:2005bu} that relatively rare phenomena in the mSUGRA
parameter space become much more `mainstream' in NUHM models, and
hence they make important cases for study, given that they are not
excluded experimentally.

Our recent work has involved the use of Markov Chain methods to
generalise the parameter spaces one can constrain using ATLAS data
\cite{Lester:2005je}. In the process, we have found that NUHM models
are an interesting testing ground for new effects, and the purpose of
this note is to highlight a particular observation related to cascade
decays. For our particular model choice, decays featuring chains of
successive two body decays are not present, and yet it is possible to
observe decay chains involving a combination of two and three body
decays. Thus, it should still be possible to measure masses by the
standard method of searching for kinematic endpoints, but we will
first need to derive expressions for their expected position. We note
that this introduces an extra layer of ambiguity, since most previous
studies have implicitly assumed that endpoints are due to two-body
decay chains. Cascade decays featuring the three body decay mode could
also occur in, for example, mSUGRA models\footnote{A suitable mSUGRA
model for study would be obtained by taking the parameters of the NUHM
benchmark point studied here and setting the GUT scale Higgs masses to
the universal scalar mass $m_0$.}, though to the best of our knowledge
they have not been studied before.

Section \ref{model} summarises our particular choice of NUHM model, reviewing both the mass spectrum and the relevant decay channels. We derive the three body endpoint positions for a general decay in section \ref{endpoints}, before going on to apply these to our NUHM model in section 4. Finally, section \ref{discussion} discusses the prospects for more detailed analysis of the NUHM model, before section \ref{conclusions} gives our final conclusions. 

\section{Selection of NUHM model}\label{model}
The NUHM parameter space is related to that of mSUGRA by the addition
of two extra parameters that express the non-universality of the two
MSSM Higgs doublets. These can be specified at the GUT scale as the
masses $m_{H_u}^2$ and $m_{H_d}^2$, or alternatively the conditions of
electroweak symmetry breaking allow one to trade these for the weak
scale parameters $\mu$ and $m_A$. In selecting a model for study, we
found the benchmark model $\gamma$ in \cite{DeRoeck:2005bw} to be
particularly interesting. The two body decay modes of the $\chi_2^0$
are not allowed, and hence one will not observe the characteristic two
body endpoints seen in a variety of mSUGRA parameter space but,
rather, will have to develop other strategies for
analysis. Furthermore, it is compatible with all current experimental
constraints arising from, e.g. WMAP and limits on the branching ratios
of rare decays.

\DOUBLEFIGURE[t]{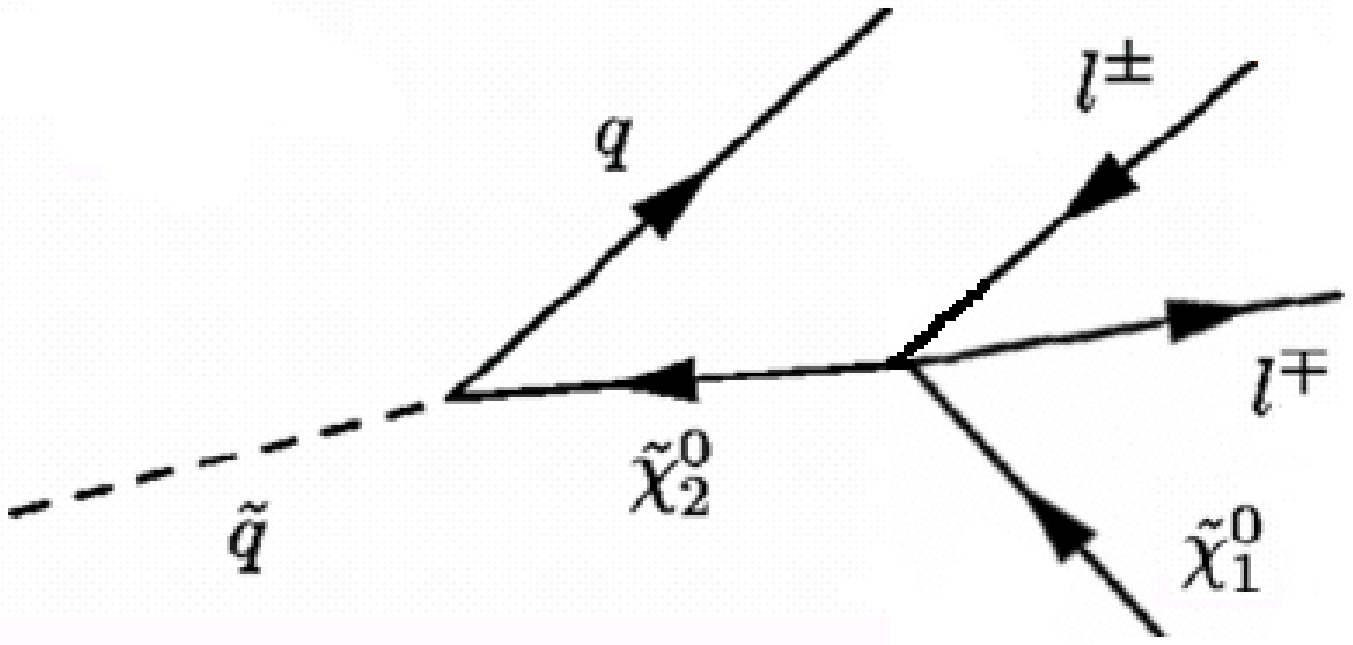,width=2.6in}{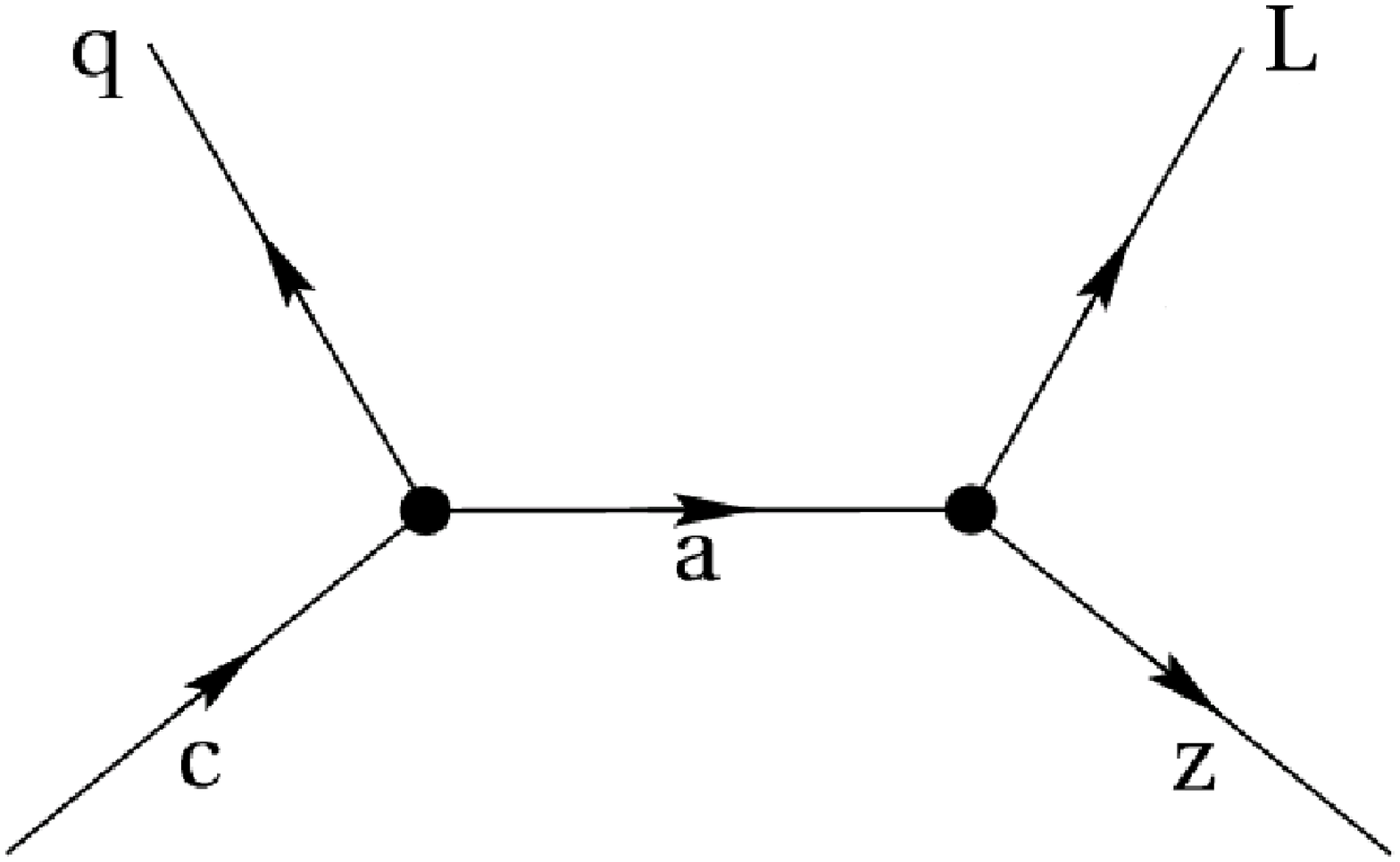,width=2.6in}{This
cascade decay chain, \label{decaypicture}including a three body decay,
will give rise to kinematic endpoints.}{The same decay
\label{decay2}as in figure 1, with the two leptons treated as a single
object $L$.}

The $\gamma$ benchmark point is specified as follows:
\begin{center} $m_0=328$ GeV, $m_{1/2}=247$ GeV \

 $tan\beta = 20$, $A_0=0$, $|\mu| = 325$ GeV\ and $m_A = 240$ GeV
\end{center}
with a top quark mass of 178 GeV, and $\mu$ greater than zero.

We have used \tt ISAJET 7.72 \rm \cite{Isajet} to generate the mass
spectrum and decay information for the point, and we summarise the
results in tables \ref{tablemasses}, and tables \ref{tabledecayschi}
and \ref{tabledecays} respectively. In addition, we used \tt HERWIG
6.5 \rm \cite{Moretti:2002eu} to estimate a total SUSY production
cross-section of 33 pb at this point. This differs from that given in
reference \cite{DeRoeck:2005bw}, though is consistent with the fact
that Herwig calculation is only performed to leading order, whereas
\cite{DeRoeck:2005bw} quotes a next to leading order result.

\DOUBLETABLE[t]
{\begin{tabular}{|c|c|}
\hline
Particle&Mass/GeV\\
\hline
$\tilde{\chi}_1^0$&95\\
$\tilde{\chi}_2^0$&179\\
$\tilde{\chi}_3^0$&332\\
$\tilde{\chi}_4^0$&353\\
$\tilde{\chi}_1^{\pm}$&179\\
$\tilde{\chi}_2^{\pm}$&353\\
\hline
$\tilde{e}_L$&377\\
$\tilde{e}_R$&329\\
$\tilde{\nu_e}$&368\\
$\tilde{\tau}_1$&315\\
$\tilde{\tau}_2$&378\\
$\tilde{\nu_{\tau}}$&365\\
\hline
$\tilde{g}$&615\\
$\tilde{u}_L$&631\\
$\tilde{u}_R$&624\\
$\tilde{d}_L$&636\\
$\tilde{d}_R$&617\\
$\tilde{b}_1$&560\\
$\tilde{b}_2$&604\\
$\tilde{t}_1$&455\\
$\tilde{t}_2$&614\\
\hline
$h^0$&116\\
$H^0$&242\\
$A^0$&240\\
$H^{\pm}$&255\\
\hline
\end{tabular}}
{\begin{tabular}{|l|r|}
\hline
Decay Mode&BR\\
\hline 
$\tilde{\chi}_2^0\rightarrow\tilde{\chi}_1^0q\bar{q}$&62\%\\
$\tilde{\chi}_2^0\rightarrow\tilde{\chi}_1^0b\bar{b}$&19\%\\
$\tilde{\chi}_2^0\rightarrow\tilde{\chi}_1^0l^+l^-$&3.5\%\\
$\tilde{\chi}_2^0\rightarrow\tilde{\chi}_1^0\tau^+\tau^-$&2.7\%\\
$\tilde{\chi}_2^0\rightarrow\tilde{\chi}_1^0\nu_{l}\bar{\nu_{l}}$&7.9\%\\
$\tilde{\chi}_2^0\rightarrow\tilde{\chi}_1^0\nu_{\tau}\bar{\nu_{\tau}}$&3.9\%\\
&\\
$\tilde{\chi}_3^0\rightarrow\tilde{\chi}_1^{\pm}W$&62\%\\
$\tilde{\chi}_3^0\rightarrow\tilde{\chi}_1^{0}Z$&14\%\\
$\tilde{\chi}_3^0\rightarrow\tilde{\chi}_2^{0}Z$&21\%\\
$\tilde{\chi}_3^0\rightarrow\tilde{\chi}_1^{0}h$&2.7\%\\
&\\
$\tilde{\chi}_4^0\rightarrow\tilde{\chi}_1^{\pm}W$&67\%\\
$\tilde{\chi}_4^0\rightarrow\tilde{\chi}_1^{0}Z$&3.4\%\\
$\tilde{\chi}_4^0\rightarrow\tilde{\chi}_2^{0}Z$&2.7\%\\
$\tilde{\chi}_4^0\rightarrow\tilde{\chi}_1^{0}h$&9.1\%\\
$\tilde{\chi}_4^0\rightarrow\tilde{\chi}_1^{0}H$&1.2\%\\
$\tilde{\chi}_4^0\rightarrow\tilde{\chi}_2^{0}h$&16.5\%\\
&\\
$\tilde{\chi}_1^{\pm}\rightarrow\tilde{\chi}_1^0W^{\pm}$&99\%\\
&\\
$\tilde{\chi}_2^{\pm}\rightarrow\tilde{\chi}_1^0W^{\pm}$&9.7\%\\
$\tilde{\chi}_2^{\pm}\rightarrow\tilde{\chi}_2^0W^{\pm}$&39\%\\
$\tilde{\chi}_2^{\pm}\rightarrow\tilde{\chi}_1^{\pm}Z$&30\%\\
$\tilde{\chi}_2^{\pm}\rightarrow\tilde{\chi}_1^{\pm}h$&20\%\\
\hline
\end{tabular}}
{The mass spectrum \label{tablemasses} of the NUHM point defined in the text, as given by \tt ISAJET 7.72 \rm.}
{The dominant chargino and neutralino decay processes \label{tabledecayschi} at the NUHM point defined in the text, as given by \tt ISAJET 7.72 \rm, where $q$ denotes a quark from the first two generations, and $l$ is a lepton from the first two generations.}

The most relevant part of the decay table concerns the decay modes of the $\tilde{\chi}^0_2$, and we see in table \ref{tabledecayschi} that we do not obtain two body decays to sleptons, but rather have three body decays to quarks or leptons. Given that we have appreciable branching fractions for squark decays featuring $\tilde{\chi}^0_2$'s we will obtain decay chains of the form shown in figure \ref{decaypicture}, and thus we should be able to observe kinematic endpoints in lepton-jet invariant mass distributions using a similar method to that which has been previously documented for chains of successive two body decays. Each maximum will occur at a position given by a function of the three sparticle masses in the decay chain. Note that although the branching ratio for the decay $\tilde{\chi}_2^0\rightarrow\tilde{\chi}_1^0l^+l^-$ is small, the large SUSY production cross-section will guarantee a reasonable sample of events (approximately 3000 events for an initial ATLAS sample of $30 \mbox{ fb}^{-1}$). 

\TABLE{\begin{tabular}{|l|r|l|r|l|r|}
\hline
Decay Mode&BR&Decay Mode&BR&Decay Mode&BR\\
\hline 
$(\tilde{u}_L,\tilde{c}_L)\rightarrow\tilde{\chi}^0_2q$&30\%&$(\tilde{u}_R,\tilde{c}_R)\rightarrow\tilde{\chi}^0_1q$&96\%&$\tilde{g}\rightarrow\tilde{b}_1b$&81\%\\
$(\tilde{u}_L,\tilde{c}_L)\rightarrow\tilde{\chi}^0_4q$&2\%&$(\tilde{u}_R,\tilde{c}_R)\rightarrow\tilde{\chi}^0_2q$&1\%&$\tilde{g}\rightarrow\tilde{b}_2b$&4\%\\
$(\tilde{u}_L,\tilde{c}_L)\rightarrow\tilde{g}q$&1.5\%&$(\tilde{u}_R,\tilde{c}_R)\rightarrow\tilde{g}q$&2.3\%&$\tilde{g}\rightarrow\tilde{\chi}_1^{\pm}q\bar{q}$&6.8\%\\
$(\tilde{u}_L,\tilde{c}_L)\rightarrow\tilde{\chi}^+_1q$&63\%&&&$\tilde{g}\rightarrow\tilde{\chi}_1^{0}q\bar{q}$&2.2\%\\
$(\tilde{u}_L,\tilde{c}_L)\rightarrow\tilde{\chi}^+_2q$&2.5\%&&&$\tilde{g}\rightarrow\tilde{\chi}_2^{0}q\bar{q}$&3.4\%\\
&&&&&\\
$(\tilde{d}_L,\tilde{s}_L)\rightarrow\tilde{\chi}^0_1q$&2.1\%&$(\tilde{d}_R,\tilde{s}_R)\rightarrow\tilde{\chi}^0_1q$&98\%&&\\
$(\tilde{d}_L,\tilde{s}_L)\rightarrow\tilde{\chi}^0_2q$&30\%&$(\tilde{d}_R,\tilde{s}_R)\rightarrow\tilde{\chi}^0_2q$&1\%&&\\
$(\tilde{d}_L,\tilde{s}_L)\rightarrow\tilde{\chi}^0_4q$&2.7\%&&&&\\
$(\tilde{d}_L,\tilde{s}_L)\rightarrow\tilde{\chi}^-_1q$&56\%&&&&\\
$(\tilde{d}_L,\tilde{s}_L)\rightarrow\tilde{\chi}^-_2q$&8\%&&&&\\
&&&&&\\
$\tilde{b}_1\rightarrow\tilde{\chi}_1^0b$&3.6\%&$\tilde{t}_1\rightarrow\tilde{\chi}_1^0t$&17\%&&\\
$\tilde{b}_1\rightarrow\tilde{\chi}_2^0b$&26\%&$\tilde{t}_1\rightarrow\tilde{\chi}_2^0t$&13\%&&\\
$\tilde{b}_1\rightarrow\tilde{\chi}_3^0b$&2.2\%&$\tilde{t}_1\rightarrow\tilde{\chi}_1^+b$&50\%&&\\
$\tilde{b}_1\rightarrow\tilde{\chi}_4^0b$&2.3\%&$\tilde{t}_1\rightarrow\tilde{\chi}_2^+b$&20\%&&\\
$\tilde{b}_1\rightarrow\tilde{\chi}_1^-t$&36\%&&&&\\
$\tilde{b}_1\rightarrow\tilde{\chi}_2^-t$&26\%&&&&\\
$\tilde{b}_1\rightarrow\tilde{t}_1W$&3.8\%&$\tilde{t}_2\rightarrow\tilde{t}_1h$&3.6\%&&\\
$\tilde{b}_2\rightarrow\tilde{\chi}_1^0b$&13\%&$\tilde{t}_2\rightarrow\tilde{\chi}_1^0t$&1.8\%&&\\
$\tilde{b}_2\rightarrow\tilde{\chi}_2^0b$&2.4\%&$\tilde{t}_2\rightarrow\tilde{\chi}_2^0t$&8.5\%&&\\
$\tilde{b}_2\rightarrow\tilde{\chi}_3^0b$&13\%&$\tilde{t}_2\rightarrow\tilde{\chi}_3^0t$&9.5\%&&\\
$\tilde{b}_2\rightarrow\tilde{\chi}_4^0b$&14\%&$\tilde{t}_2\rightarrow\tilde{\chi}_4^0t$&27\%&&\\
$\tilde{b}_2\rightarrow\tilde{\chi}_1^-t$&3.2\%&$\tilde{t}_2\rightarrow\tilde{\chi}_1^+b$&22\%&&\\
$\tilde{b}_2\rightarrow\tilde{\chi}_2^-t$&46\%&$\tilde{t}_2\rightarrow\tilde{\chi}_2^+b$&21\%&&\\
$\tilde{b}_2\rightarrow\tilde{t}_1W$&8.2\%&$\tilde{t}_2\rightarrow\tilde{t}_1Z$&7\%&&\\
&&&&&\\
$\tilde{l}_L\rightarrow\tilde{\chi}_1^0l$&12\%&$\tilde{l}_R\rightarrow\tilde{\chi}_1^0l$&99\%&$\tilde{\nu}_l\rightarrow\tilde{\chi}_1^0\nu_l$&17\%\\
$\tilde{l}_L\rightarrow\tilde{\chi}_2^0l$&33\%&&&$\tilde{\nu}_l\rightarrow\tilde{\chi}_2^0\nu_l$&24\%\\
$\tilde{l}_L\rightarrow\tilde{\chi}_1^-\nu_e$&54\%&&&$\tilde{\nu}_l\rightarrow\tilde{\chi}_1^+l$&59\%\\
$\tilde{\tau}_1\rightarrow\tilde{\chi}_1^0\tau$&81\%&$\tilde{\tau}_2\rightarrow\tilde{\chi}_1^0\tau$&16\%&$\tilde{\nu}_{\tau}\rightarrow\tilde{\chi}_1^0\nu_{\tau}$&17\%\\
$\tilde{\tau}_1\rightarrow\tilde{\chi}_2^0\tau$&6.9\%&$\tilde{\tau}_2\rightarrow\tilde{\chi}_2^0\tau$&32\%&$\tilde{\nu}_{\tau}\rightarrow\tilde{\chi}_2^0\nu_{\tau}$&24\%\\
$\tilde{\tau}_1\rightarrow\tilde{\chi}_1^-\nu_{\tau}$&12\%&$\tilde{\tau}_2\rightarrow\tilde{\chi}_1^-\nu_{\tau}$&50\%&$\tilde{\nu}_{\tau}\rightarrow\tilde{\chi}_1^+\tau$&60\%\\
\hline
\end{tabular}
\caption{The dominant sfermion decay processes at the NUHM point defined in the text, as given by \tt ISAJET 7.72 \rm, where $q$ denotes a quark from the first two generations, and $l$ is a lepton from the first two generations.\label{tabledecays}} 
}

\section{Kinematic endpoint derivation}\label{endpoints} 
\label{sec:sdfhdjkgjytiretrkfjb}
\subsection{Introduction}
The R-parity conservation referred to in section \ref{intro} has important implications for collider experiments: sparticles must be pair produced, and the LSP is stable. Thus, if R-parity is indeed conserved, each SUSY event at the LHC will have two sparticle decay chains, and the escaping LSPs will make it difficult to fully reconstruct events. It is, however, possible to construct distributions that are sensitive to sparticle masses. 

In this paper we consider the decay $\tilde{q} \to q \tilde{\chi}_2^0$ followed by $\tilde{\chi}_2^0 \to l^+ l^- \tilde{\chi}_1^0$ as shown in figure \ref{decaypicture}. Such decays are fairly easy to select given that one can look for events with opposite-sign-same-flavour (OSSF) leptons, combined with the missing energy from the undetected neutralinos\footnote{Note that this does not preclude the possibility of selecting the usual two body cascade process, and we consider ways to resolve this ambiguity in section \ref{discussion}.}. By taking different combinations of the visible decay products, one can form various invariant masses; $m_{ll}$, $m_{llq}$, $m_{lq}^{high}$ and $m_{lq}^{low}$, where $m_{lq}^{high}$ is the higher of the two $m_{lq}$ invariant masses that can be formed in the event, and $m_{lq}^{low}$ is the lower. These will have maxima resulting from kinematic limits, whose position is given by a function of $m_{\tilde{q}}$, $m_{\tilde{\chi}_2^0}$ and $m_{\tilde{\chi}_1^0}$, and we derive these for each case below.

In the following derivations, we will use bold type for three momenta, and will denote four vector quantities by explicitly showing Lorentz indices. In addition, we will introduce a convention for representing squared masses by a non-bold character (e.g. $c = m_{\tilde{q}}^2$). 

\subsection{$m_{ll}$ endpoint}

The endpoint of the $m_{ll}$ distribution results from the three body decay of the $\tilde{\chi}_2^0$, and is given trivially by the mass difference between the $\tilde{\chi}_2^0$ and the $\tilde{\chi}_1^0$:
\begin{equation}\label{lledge}
(m_{ll}^2)^{\mbox{\tiny max}}=(m_{\tilde{\chi}^0_2}-m_{\tilde{\chi}^0_1})^2.
\end{equation}
\subsection{$m_{llq}$ endpoint and threshold}
In calculating the $m_{llq}$ endpoint, we follow the method given in appendix E of \cite{Lester:2001zx} and treat the decay as shown in figure \ref{decay2}, where we have combined the two leptons into a single object $L$, with an invariant mass given by $m_{ll}$. We know from the dilepton invariant mass that $m_{ll}\equiv m_{L}$ must lie within a specific range:
\begin{equation}\label{range}
m_L = \lambda (m_a-m_z) ,\lambda \in [0,1]
\end{equation}
If we look at the decay of figure \ref{decay2} in the rest frame of $a$, we can conserve four momentum to obtain the following expressions for the three momenta of $q$ and $L$:
\begin{equation}\label{L}
\textbf{L}^2 = \textbf{z}^2  = [m_L^2,m_a^2,m_z^2]
\end{equation}
\begin{equation}\label{q}
\textbf{q}^2 = \textbf{c}^2  = [0,m_a^2,m_c^2]
\end{equation}
where
\begin{equation}
[x,y,z]\equiv \frac{x^2+y^2+z^2-2(xy+xz+yz)}{4y}
\end{equation}
and we have treated the quark as massless. 

Taking $q$ to be massless, the invariant mass of $q$ and $L$ is in
general given by
\begin{eqnarray}
m_{qL}^2&=&g_{\mu\nu}(L^{\mu}+q^{\mu})(L^{\nu}+q^{\nu})\\
&=&m_L^2+2|\textbf{q}|(E_L-|\textbf{L}|\cos \theta)\label{invmass}
\end{eqnarray}
in which $\theta$ is the angle between $L$ and $q$ in the rest
frame of $a$, the intermediate particle.  The maximum will occur when
$\cos \theta$ is equal to -1, and hence $L$ and $q$ are back to back in
the $a$ rest frame. Combining this with our knowledge of $|{\bf L}|$
and $|{\bf q}|$ from equations \ref{L} and \ref{q}, we obtain the
expression for the endpoint of the $m_{llq}$ distribution in terms of
$m_L$:
\begin{equation}\label{llqgeneral}
(m_{llq})^2=L_m + \frac{(c-a)}{2a} \left[ L_m-(z-a) + \sqrt{((a+z)-L_m)^2-4az} \right]
\end{equation}
where $L_m = m_L^2$, $c = m_c^2$, $a = m_a^2$ and $z = m_z^2$. $L_m$ can take any value in the range specified by equation \ref{range}, and we now need to maximise equation \ref{llqgeneral} by considering separately the cases where $\lambda = 0$, $0 < \lambda < 1$ and $\lambda = 1$. After doing this, we obtain two possible expressions for the maximum of the $m_{llq}$ distribution:
\begin{equation}\label{llqedge}
(m_{llq}^2)^{\mbox{\tiny{max}}}=\left\{ \begin{array}{l}
                            (m_{\tilde{q}}-m_{\tilde{\chi}^0_1})^2 \mbox{ if } m_{\tilde{\chi}^0_2}^2 > m_{\tilde{q}}m_{\tilde{\chi}^0_1} \\
                            (m_{\tilde{q}}^2-m_{\tilde{\chi}^0_2}^2)(m_{\tilde{\chi}^0_2}^2-m_{\tilde{\chi}^0_1}^2)/m_{\tilde{\chi}^0_2}^2 \mbox { otherwise.}
                            \end{array} \right.
\end{equation}

In addition to finding an edge in the $m_{llq}$ distribution, one can
observe a threshold. Equation \ref{invmass} has a minimum when $\cos
\theta$ is equal to 1, in which case one obtains a minimum value of
$m_{llq}$:
\begin{equation}
(m_{llq}^{min})^2=L_m + \frac{(c-a)}{2a} \left[ L_m-(z-a) -
    \sqrt{((a+z)-L_m)^2-4az} \right] \label{eq:genllqthresh}
\end{equation}
If $L_m$ lies at the lower end of its allowed range, then we have $L_m = m_{llq}^{min} = 0$. However, we can raise the minimum value of $m_{llq}$ by looking at the subset of events for which $L_m$ is greater than some arbitrary cut value. This will then give us an observable threshold in the $m_{llq}$ distribution.
\subsection{$m_{lq}^{high}$ and $m_{lq}^{low}$ endpoints}
In the case of the $m_{llq}$ endpoint, we showed that there are in fact two expressions, each of which applies in a specific region of mass space. In anticipation of this, we used a general method to avoid missing one of the solutions. 

The $m_{lq}^{high}$ endpoint is conceptually much easier, however, as we only have to maximise the invariant mass that we can make from one of the leptons. The two sequential decays are shown in figure \ref{boost}, where we show the effect of a boost from the $\tilde{q}$ rest frame to the $\tilde{\chi}_2^0$ rest frame. Any maximum in the $m_{lq}$ invariant mass must arise from having the relevant lepton (the `interesting lepton') back to back with the quark in the $\tilde{\chi}^0_2$ rest frame. We can thus consider three extreme cases for the configuration of the leptons and $\tilde{\chi}_1^0$ in the $\tilde{\chi}_2^0$ rest frame:
\begin{figure}
\centerline{
\includegraphics[width=2.4in]{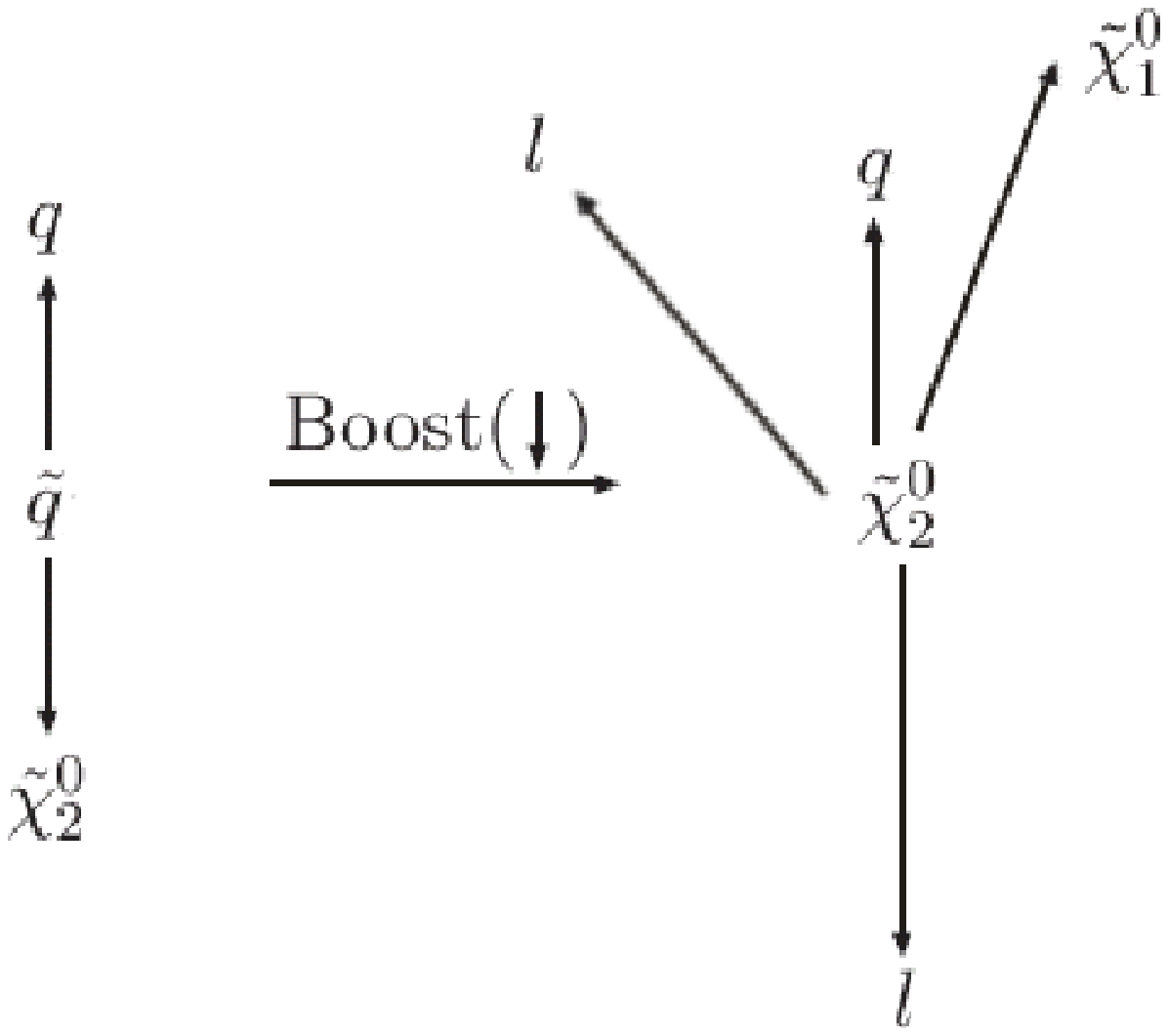}
}
\caption{The two rest frames involved in the squark cascade decay.\label{boost}}
\end{figure}
\begin{enumerate}
\item
The $\tilde{\chi}^0_1$ is produced at rest, and the two leptons are back to back with one of them anti-parallel to the quark.
\item
One of the leptons is produced at rest, and so the $\tilde{\chi}^0_1$ is produced back to back with the other lepton, with the interesting lepton being anti-parallel to the quark.
\item
None of the particles from the three body decay is produced at rest, in which case we will get the highest invariant mass by having the interesting lepton emerging anti-parallel to the quark, and the other two particles travelling in the same direction as the quark.
\end{enumerate}
Obtaining the $m_{lq}$ endpoint is simply a case of working out which of these gives the highest invariant mass. A short calculation gives us:
\begin{equation}\label{lqhighedge}
(m_{lq}^2)^{\mbox{\tiny max}}_{\mbox{\tiny high}}=\frac{(m_{\tilde{q}}^2-m_{\tilde{\chi}^0_2}^2)(m_{\tilde{\chi}^0_2}^2-m_{\tilde{\chi}^0_1}^2)}{m_{\tilde{\chi}^0_2}^2}
\end{equation}

The $m_{lq}^{low}$ endpoint is harder to obtain than the $m_{lq}^{high}$ endpoint, given that we want the maximum value of the smallest $m_{lq}$ invariant mass it is possible to form in an event. We apply a similar approach to that used in the previous subsection, and visualise the decay configuration that will give us the maximum before working out the endpoint. 

In this case, we have proved that there is a local $m_{lq}^{low}$
maximum when the two leptons are produced parallel in the
$\tilde{\chi}^0_2$ rest frame, travelling anti-parallel to the
$\tilde{\chi}^0_1$ (which therefore travels parallel to the
quark). Note that this does not exclude the possibility of other local
maxima, but numerical simulation has not revealed any. We therefore take
this configuration to give the global maximum of
$m_{lq}^{low}$.\footnote{Other local maxima, were any to exist, would
occur in configurations in which $m_{lq}^{low}$ were equal to
$m_{lq}^{high}$ but in which the moduli of the three momenta of the
leptons were unequal (but coplanar with the quark) in the rest frame of the heavier neutralino.}

A short calculation gives:
\begin{equation}\label{lqlowedge}
(m_{lq}^2)^{\mbox{\tiny max}}_{\mbox{\tiny low}}=\frac{(m_{\tilde{q}}^2-m_{\tilde{\chi}^0_2}^2)(m_{\tilde{\chi}^0_2}^2-m_{\tilde{\chi}^0_1}^2)}{2m_{\tilde{\chi}^0_2}^2}
\end{equation}

\subsection{Summary}
Having obtained endpoints for the $m_{ll}$, $m_{llq}$, $m_{lq}^{high}$ and $m_{lq}^{low}$ distributions, we see that the expressions are very similar. The ratio of the $m_{lq}^{high}$ and $m_{lq}^{low}$ endpoint positions will always be $\sqrt{2}$, and, in a particular mass region, the $m_{llq}$ endpoint is coincident with the $m_{lq}^{high}$ endpoint. This ultimately means that there may not be enough information to precisely determine the mass differences involved in the decay chain, a point that will be discussed further in section \ref{discussion}.

\section{Observation of three body endpoints in NUHM model}
Having derived the endpoints for the process depicted in figure \ref{decaypicture}, we now discuss a concrete physics example by performing a Monte Carlo study of the NUHM model described in section \ref{model}.

\subsection{Event generation and simulation}
The mass spectrum and decay table of the NUHM point were taken from \tt ISAJET 7.72 \rm using the \tt ISAWIG \rm interface. We subsequently generated 3,300,000 signal events (corresponding to an integrated luminosity of $100 \mbox{ fb}^{-1}$), using \tt HERWIG 6.5 \rm. This implements three body decays of SUSY particles with spin correlations, with the decays of interest here being: 
\begin{enumerate}
\item{$\tilde{\chi}_2^0 \to Z \tilde{\chi}_1^0 \to l^+ l^- \tilde{\chi}_1^0$}
\item{$\tilde{\chi}_2^0 \to l \tilde{l} \to l^+ l^- \tilde{\chi}_1^0$}
\end{enumerate}
where the $Z$ and the slepton are off-shell. These generated events are then passed through the \tt ATLFAST \rm detector simulation (from the same release), whose jet cone algorithm used a cone with $\Delta R = 0.4$. Electrons, muons and jets were subject to a minimum $p_T$ cut of 5, 6 and 10 GeV respectively.

We note that the \tt ATLFAST \rm reconstruction algorithms affect the ability to reconstruct leptons and jets in close proximity, and this is potentially a source of systematic error in our endpoints observations (particularly in the threshold position). This will also occur in full simulation, though to a lesser extent. A study of these systematic effects in both the fast and full simulation is long overdue, but is sufficiently complicated to warrant a separate publication. We note in passing that this affects all previous endpoint analyses, and is not specific to that considered here.

\subsection{Selection cuts}
In order to observe clear endpoints from the cascade decay process, it is necessary to first isolate a clean sample of squark decay events. One can select events with two OSSF leptons and a large amount of missing energy, and can also exploit the fact that one expects hard jets in SUSY events. Hence, all plots that follow are subject to the following cuts:
\begin{itemize}
\item
$E_{T}^{\mbox{\tiny miss}} > 300 \mbox{ GeV;}$
\item
exactly two opposite-sign leptons with $p_T > 5 \mbox{ GeV}$ for electrons and $p_T > 5 \mbox{ GeV}$ for muons, with $|\eta| < 2.5$;
\item
at least two jets with $p_T > 150 \mbox{ GeV}$;
\end{itemize}
The $E_{T}^{\mbox{\tiny miss}}$ and jet requirements should be
sufficient to ensure that the events would be triggered by ATLAS.  For
example, these events should pass both specialised Supersymmetry
trigger (``j70+xE70'': 1 jet $>$ 70 GeV and $E_{T}^{\mbox{\tiny miss}}
> 70$\ GeV) and the inclusive missing energy trigger (``xE200'':
$E_{T}^{\mbox{\tiny miss}} > 200$\ GeV) -- see \cite{atlasTrigger2003}
and references therein.  There are of course many SM processes that
contribute to the dilepton background in any SUSY analysis, though it
has been shown that these are highly suppressed once an OSSF cut is
used in conjunction with cuts on lepton and jet $p_T$, and on missing
energy. Although there is in principle still a tail of SM events that
can contribute, it has found to be negligible in the past (see, for
example, \cite{Hinchliffe:1996iu}), and a full study of this
background is considered to be beyond the scope of this paper. We also
note that the OSSF lepton signature can be produced by SUSY processes
other than the decay of the $\tilde{\chi}_2^0$.  However, a large
fraction of such processes generate the two leptons with uncorrelated
families, and so produce an equal number of opposite-sign
opposite-flavour (OSOF) leptons.  Thus one can ``remove'' this
fraction (the majority) of the SUSY dilepton background by producing
``flavour subtracted plots'' in which one plots the combination
$e^+e^-+\mu^+\mu^--e^+\mu^--e^-\mu^+$.  Figures \ref{mll-plot} to
\ref{mlqlow} below have all been flavour subtracted.  Note that at the
end of this flavour subtraction, a small number of events from SUSY
processes producing dileptons of correlated flavour still remain.
Events of this kind may be seen in the upper tail of
figure~\ref{mllq-plot}.  Note also that events in some of the plots
below have been subjected to additional cuts (beyond the basic set
detailed above).  Where these additional cuts occur, they are
explained in the text.
\subsection{Note on endpoint positions}
It should be remembered that the formulae for endpoint positions
presented in section~\ref{sec:sdfhdjkgjytiretrkfjb} take a squark mass
as input.  In reality not all squarks have the same mass, and so
chains containing squarks with different masses will have endpoints at
slightly different positions.  This effect manifests itself as a
smearing of the endpoints in any plots of experimental or simulated
data.  Plots of this kind are shown in section~\ref{sec:sdkxfdgdgfhd}
and when indicating the positions at which endpoints are expected to
be found in these plots, we are required to choose a ``typical''
squark mass for insertion into the relevant endpoint formula.  We
choose a 610 GeV ``typical'' squark mass for this purpose, although it
should be borne in mind that the actual endpoints seen in the plots
will be somewhat smeared due the the non-degeneracy of the squark
masses contributing to them.
\subsection{Invariant mass plots}
\label{sec:sdkxfdgdgfhd}
\subsubsection{$m_{ll}$ plot}
The $m_{ll}$ distribution is shown in figure \ref{mll-plot}; 4566 events survive after cuts and background subtraction, though it is noted that the effect of the trigger which may cut more harshly on lepton $p_T$ has not been considered. Using the mass spectrum given in section \ref{model}, we expect to find an endpoint at approximately 80 GeV, and this is clearly visible. 

It is noted that the shape of the distribution is very different from the triangular shape normally encountered in the case of successive two body decays resulting from the phase space for that process, and this might prove important when attempting to distinguish three body from two body decays. This is considered further in section \ref{discussion}.
\subsubsection{$m_{llq}$ plots}
As soon as we start to form invariant masses involving quarks, it is important to consider how to select the correct quark from the cascade decay. A reasonable assumption is that the two hardest jets in the event will come from squark decay on either side of the event, and if we take the lowest of the two $m_{llq}$ invariant masses formed from the two hardest jets in the event, this should lie below the $m_{llq}$ endpoint. The distribution of this $m_{llq}$ is shown in figure \ref{mllq-plot}, and there is a visible endpoint consistent with the predicted value of approximately 490 GeV. The plot contains the same number of events as the $m_{ll}$ plot, as the cuts are the same. 

In order to obtain a further constraint on the physical model
underlying the data, we construct a threshold in the $m_{llq}$
distribution.  We follow the convention used in the study of
successive two body decays, and choose to look at the subset of events
for which $m_{ll} > m_{ll}^{max}/\sqrt{2}$.\footnote{It remains an
open question as to whether this or similar analyses would benefit
from the optimisation of the position of this cut on $m_{ll}$.}
Substituting $m_{ll}^{max}/\sqrt{2}$ in place of $L_m$ in
equation~\ref{eq:genllqthresh} gives the following threshold:
\begin{equation}
(m_{llq}^{min})^2 = \frac{(\sqrt{a}-\sqrt{z})^2}{2}+\frac{(c-a)}{4a}\left(3a - z -2\sqrt{az} - \sqrt{a^2+z^2+4\sqrt{az}(a + z) - 10az}\right)
\end{equation}
where $c=m_{\tilde{q}}^2$, $a=m_{\tilde{\chi}_2^0}^2$ and
$z=m_{\tilde{\chi}_1^0}^2$.  Traditionally (i.e.\ in chains with
successive {\em two-body} decays) this additional constraint requires,
somewhat arbitrarily, that the angular separation of the two leptons
in the rest frame of the slepton be greater than than a right angle.
For the {\em three-body} neutralino decay considered in this paper,
that geometrical interpretation is lost, but this is of no consequence
to us.

A plot of the $m_{llq}$ distribution is given in figure
\ref{llqthres}, where we note that, because we are looking for a
threshold, the highest of the two $m_{llq}$ invariant masses formed
with the two hardest jets in the event is used to make the plot. 4172
events are contained in the plot. A threshold structure of some form
is clearly observed, though it is difficult to ascertain the precise
position, as the shape of the edge is not yet a well understood
function of the sparticle masses and cut-induced `detector' effects.
To use the constraint from this edge to the full, it may be necessary
to repeat the analysis of \cite{Lester:2006yw} in the context of a
three-body final decay. The predicted value is approximately 240 GeV.

\subsubsection{$m_{lq}$ plots}
\begin{figure}
\centerline{
\includegraphics[width=2.6in]{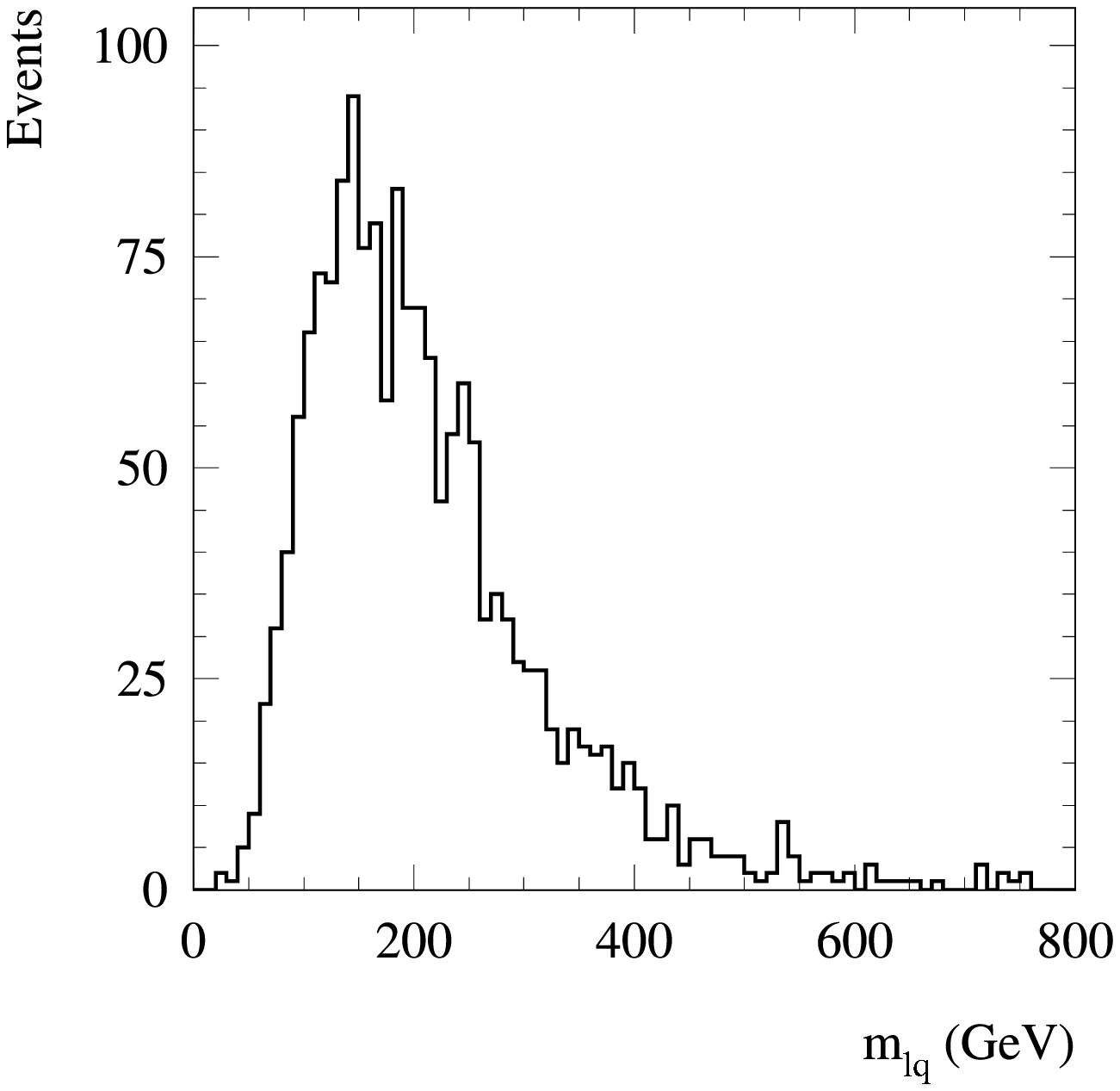}
}
\caption{The $m_{eq}^{\tiny{high}}$ distribution obtained using events with no decay of the form $\tilde{\chi}_2^0 \to \tilde{\chi}_1^0 e^+ e^-$, representing the contribution to the plot from SUSY background processes.\label{lqbackplot}}
\end{figure}

\begin{figure}
\centerline{
\includegraphics[width=2.6in]{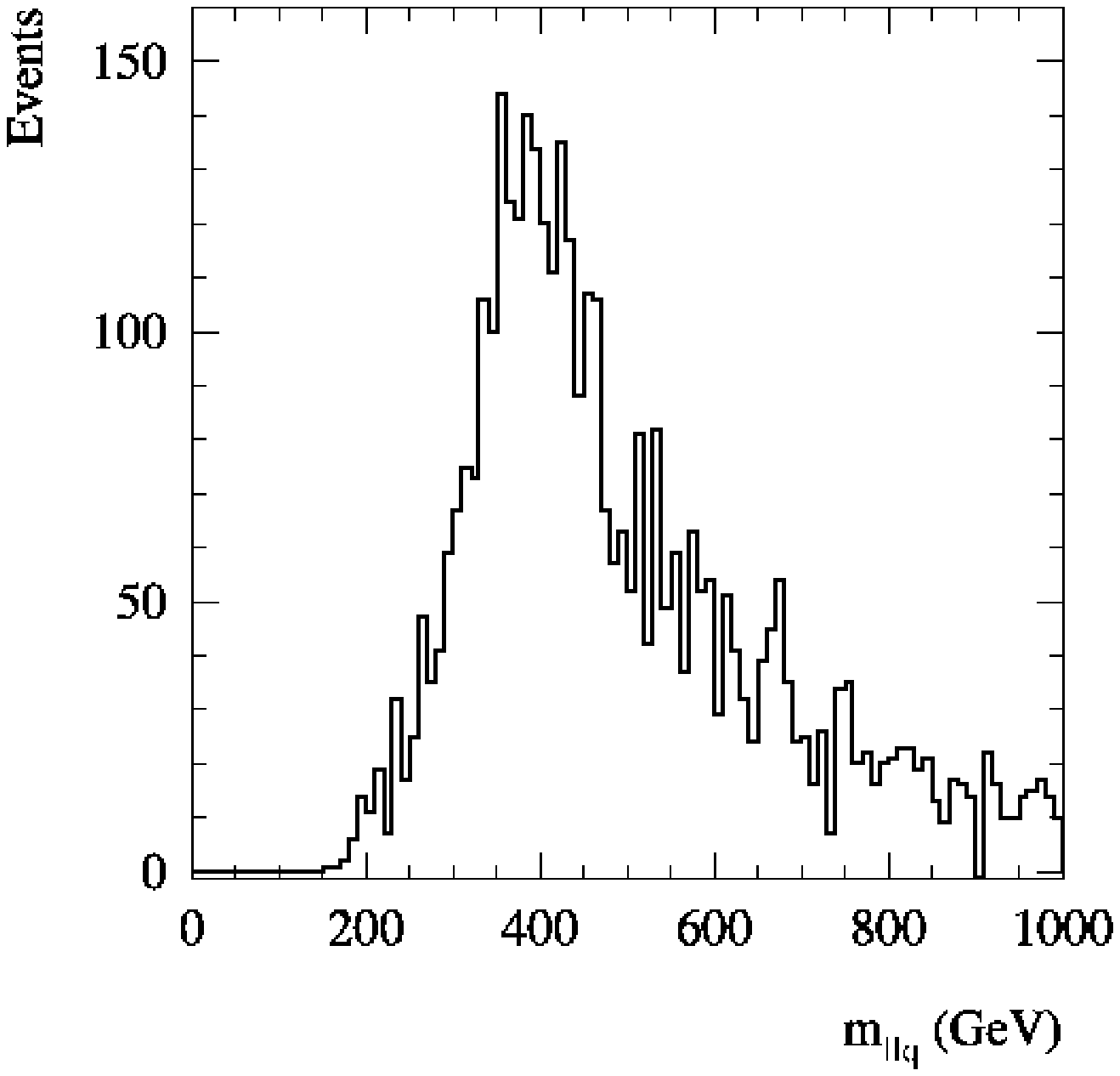}
}
\caption{The flavour subtracted $m_{llq}$ threshold plot, constructed using the highest of the $m_{llq}$ invariant masses that can be formed from the two hardest jets in each selected event.}
\label{llqthres}
\end{figure}

The $m_{lq}^{\tiny{high}}$ distribution is plotted by forming
$m_{llq}$ invariant masses with the two hardest jets in the event. The
jet from the lowest mass combination (which is our best guess for the
quark emitted in the squark cascade decay) is then used to form the
$m_{lq}$ invariant mass with each of the leptons in the events. The
maximum of these is plotted in the $m_{lq}^{\mbox{\tiny high}}$ plot
(shown in figure~\ref{mlqhigh}), where we note that we have used the
additional cut that the dilepton invariant mass in each selected event
must be less than the dilepton endpoint. 4161 events are in the
plot. There is an endpoint predicted at about 490 GeV (we are in the
mass region where it should appear at the same position as the
$m_{llq}$ endpoint) and this is consistent with the plot, though it is
difficult to identify the endpoint due the fact that the shape is
easily confused with the tail.  It is easier to see why this is the
case by looking at the distribution in a simpler context, namely one
that ignores all detector effects and which looks only at phase space
which we have implemented using a ``toy'' Monte Carlo.
In this, we also ignore the smearing coming from the spread in squark masses
which is normally present, by generating chains with a single squark mass. 
Using this toy Monte Carlo we generate plots of the distribution in the
vicinity of the edge (figures \ref{mlspace} and \ref{mlspacezoom}) and
we see that the endpoint is only approached quadratically.
Although a full
analysis of the tail would probably require full simulation (and thus
a separate study), we have attempted to determine how much is caused
by detector smearing, and how much is caused by background SUSY
processes. Figures \ref{herwig} to \ref{lqbackplot} examine the
$m_{eq}^{\tiny{high}}$ distribution, showing the Monte Carlo truth
plot, the plot obtained by selecting events on the basis of truth but
with the particles reconstructed by the \tt ATLFAST \rm detector
simulation, and finally a plot which contains only SUSY background
processes. We see that the tail does not predominantly arise from
detector smearing (which will nevertheless smear the endpoint), but
has instead a large contribution from the SUSY background. There is
also a combinatoric background related to the wrong choice of squark,
but this is harder to isolate.

\DOUBLEFIGURE[t]{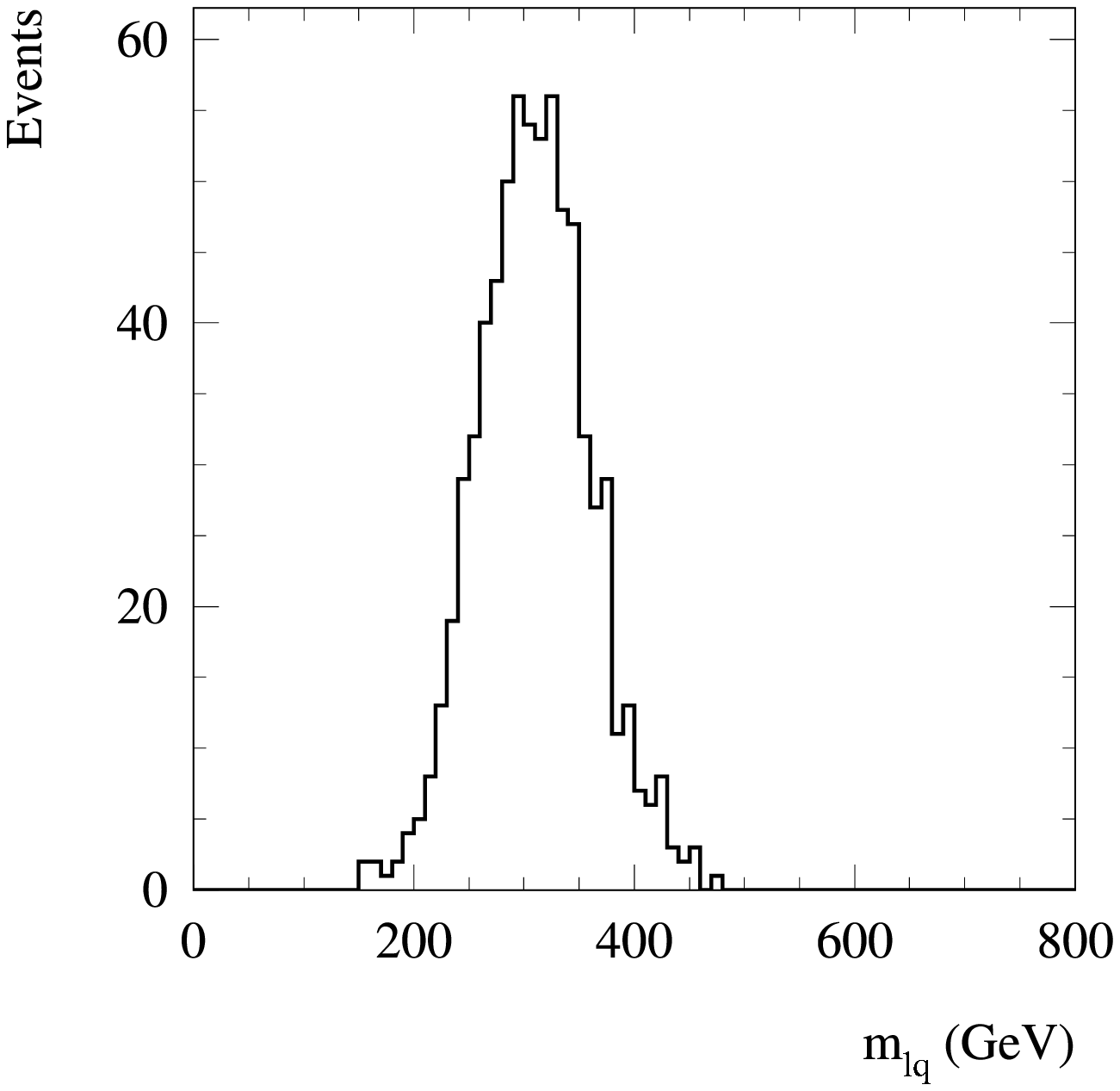,width=2.6in}{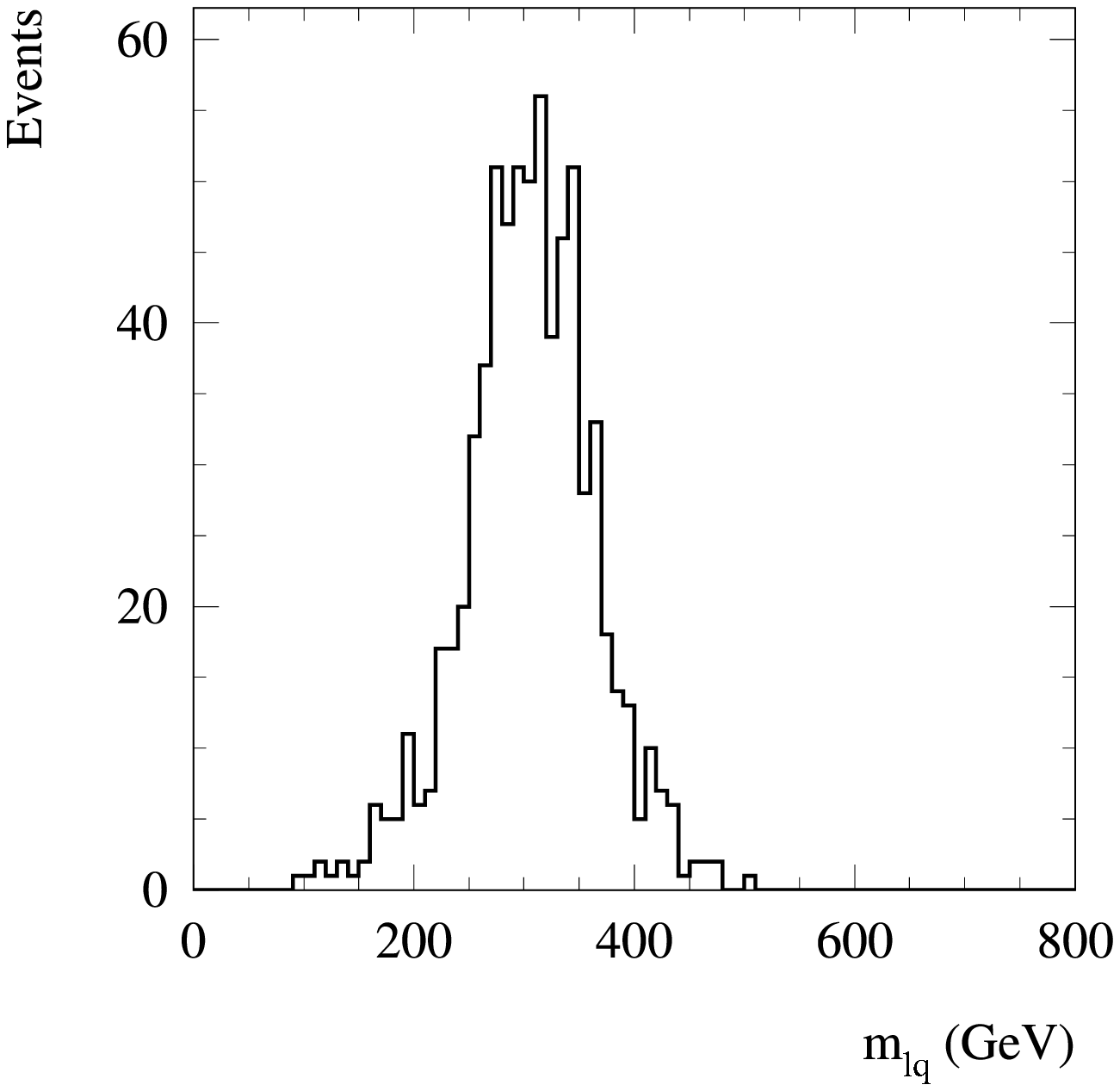,width=2.6in}{The truth distribution for the $m_{eq}^{\tiny{high}}$ invariant mass, taken from the Herwig event record as recorded in \tt ATLFAST \rm. The exhibits a clean edge with no tail.\label{herwig}}{The $m_{eq}^{\tiny{high}}$ distribution obtained by selecting events on the basis of Monte Carlo truth information, but with the electrons and jets reconstructed by \tt ATLFAST \rm. We see that the plot has a slightly higher endpoint than the truth distribution, but no significant tail.}

The $m_{lq}^{\mbox{\tiny low}}$ plot is constructed in a similar fashion to the  $m_{lq}^{\mbox{\tiny high}}$ plot, with the exception that we take the lowest of the two possible $m_{lq}$ combinations in each event. The result is shown in figure \ref{mlqlow}, where we have used an additional cut; one of the $m_{llq}$ invariant masses formed from the two hardest jets in the event must lie below the approximate observed position of the $m_{llq}$ endpoint, and one must lie above, leaving 1664 events in the plot. This removes much of the tail due to incorrect squark choice, and leaves us with a very clean endpoint at the predicted value of approximately 350 GeV.\footnote{We note that this extra cut is not possible in the case of the $m_{lq}^{high}$ endpoint, as the $m_{lq}^{high}$ distribution is highly correlated to the $m_{llq}$ distribution (the events at the endpoint are the same in both cases). Hence, performing this cut on the $m_{lq}^{high}$ distribution artificially removes any events beyond the endpoint. }

\DOUBLEFIGURE[t]{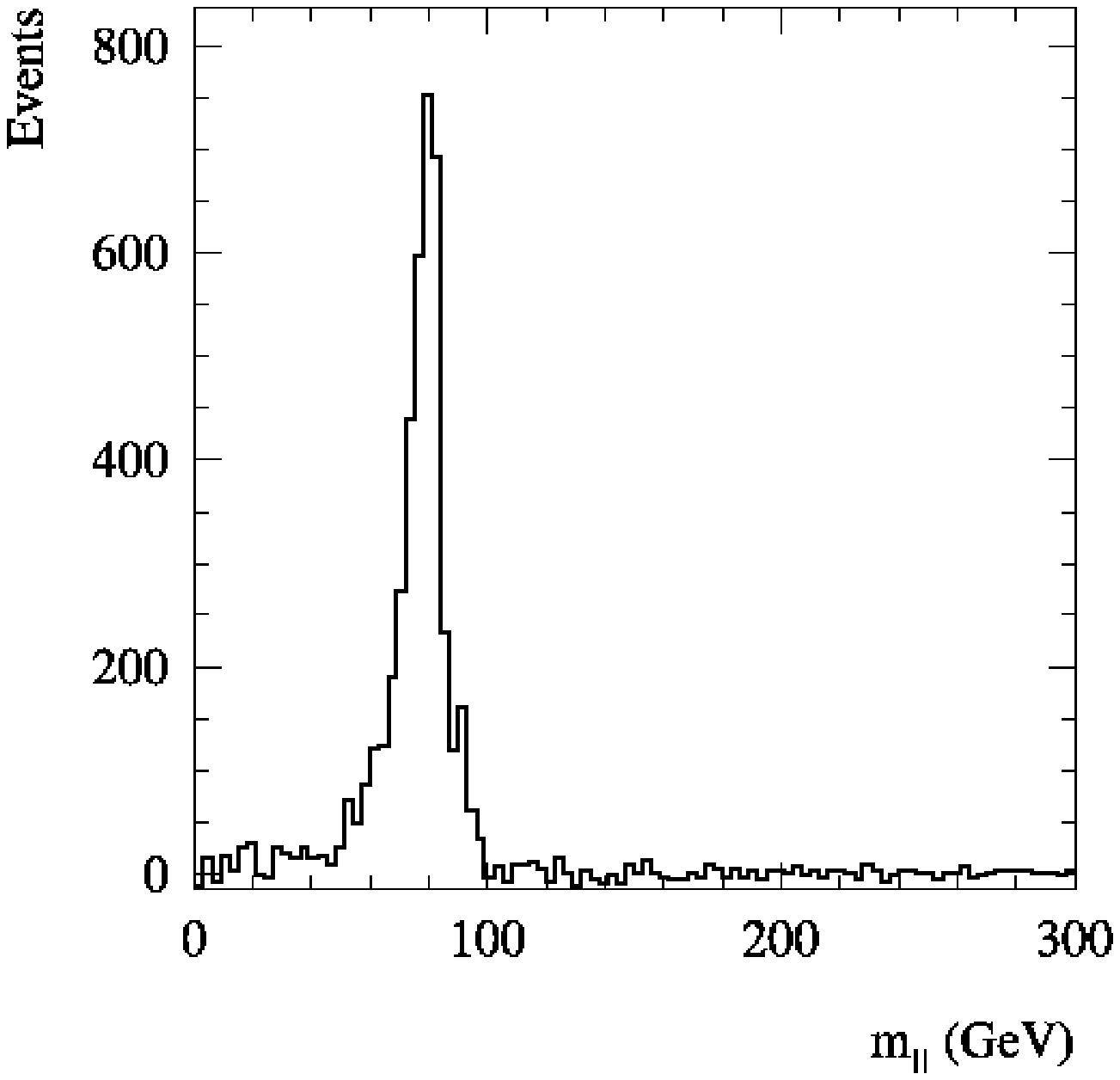,width=2.6in}{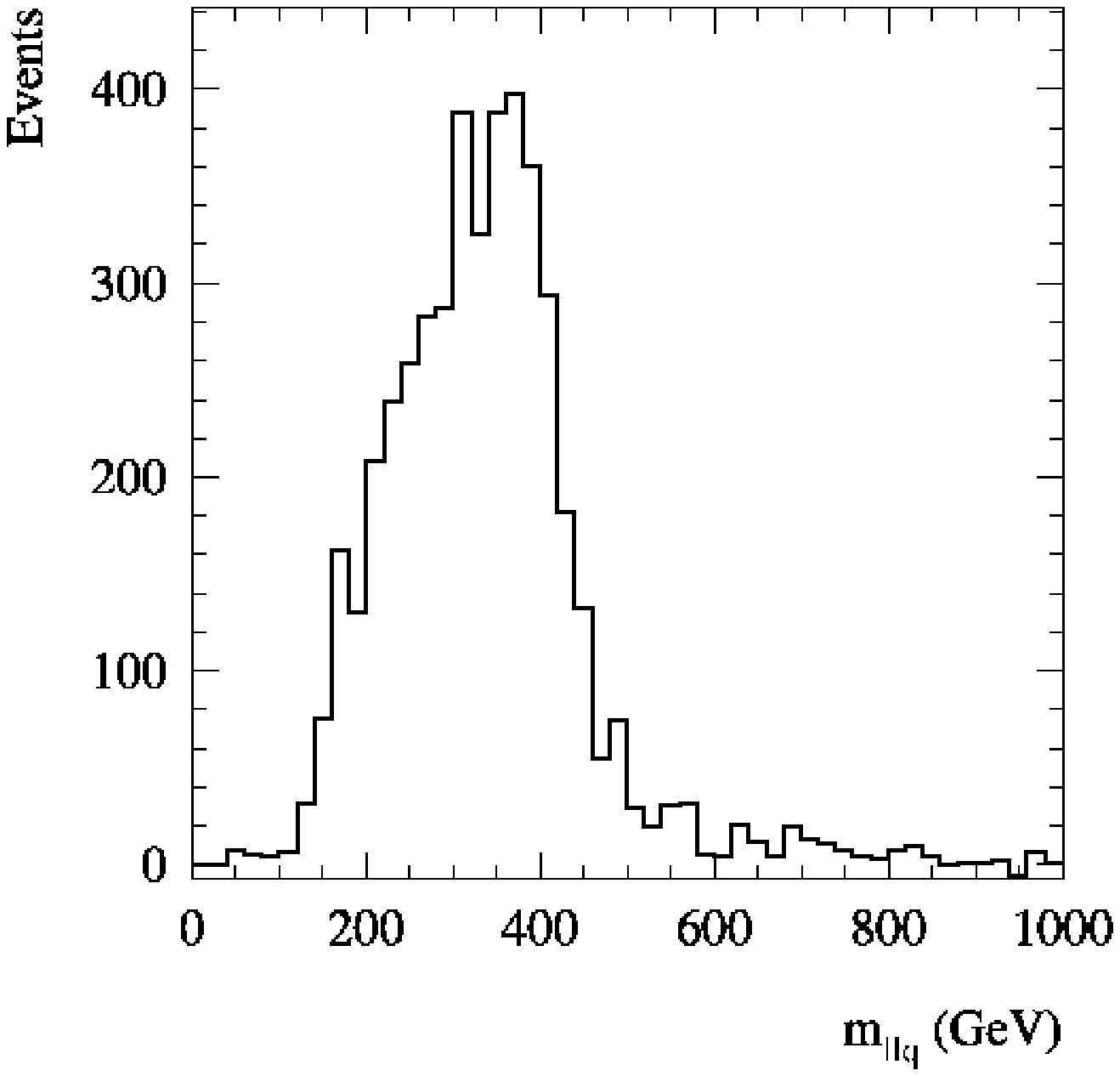,width=2.6in}{The $m_{ll}$ distribution \label{mll-plot}for the NUHM model defined in the text, with flavour subtraction.}{The flavour subtracted $m_{llq}$ distribution \label{mllq-plot} for the NUHM model defined in the text, constructed by taking the lowest $m_{llq}$ invariant mass that can be formed from the two hardest jets in the event.}

\DOUBLEFIGURE[t]{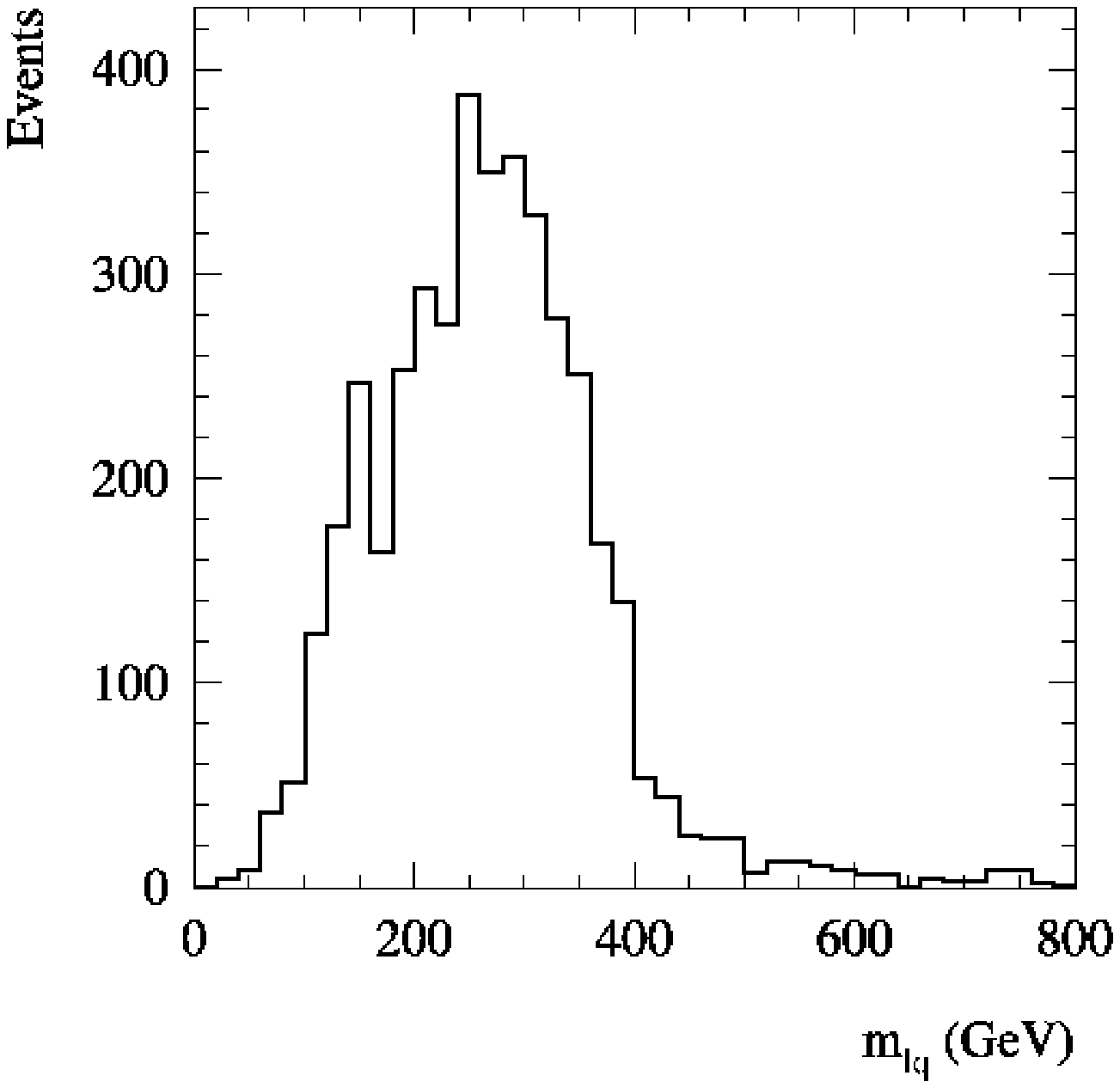,width=2.6in}{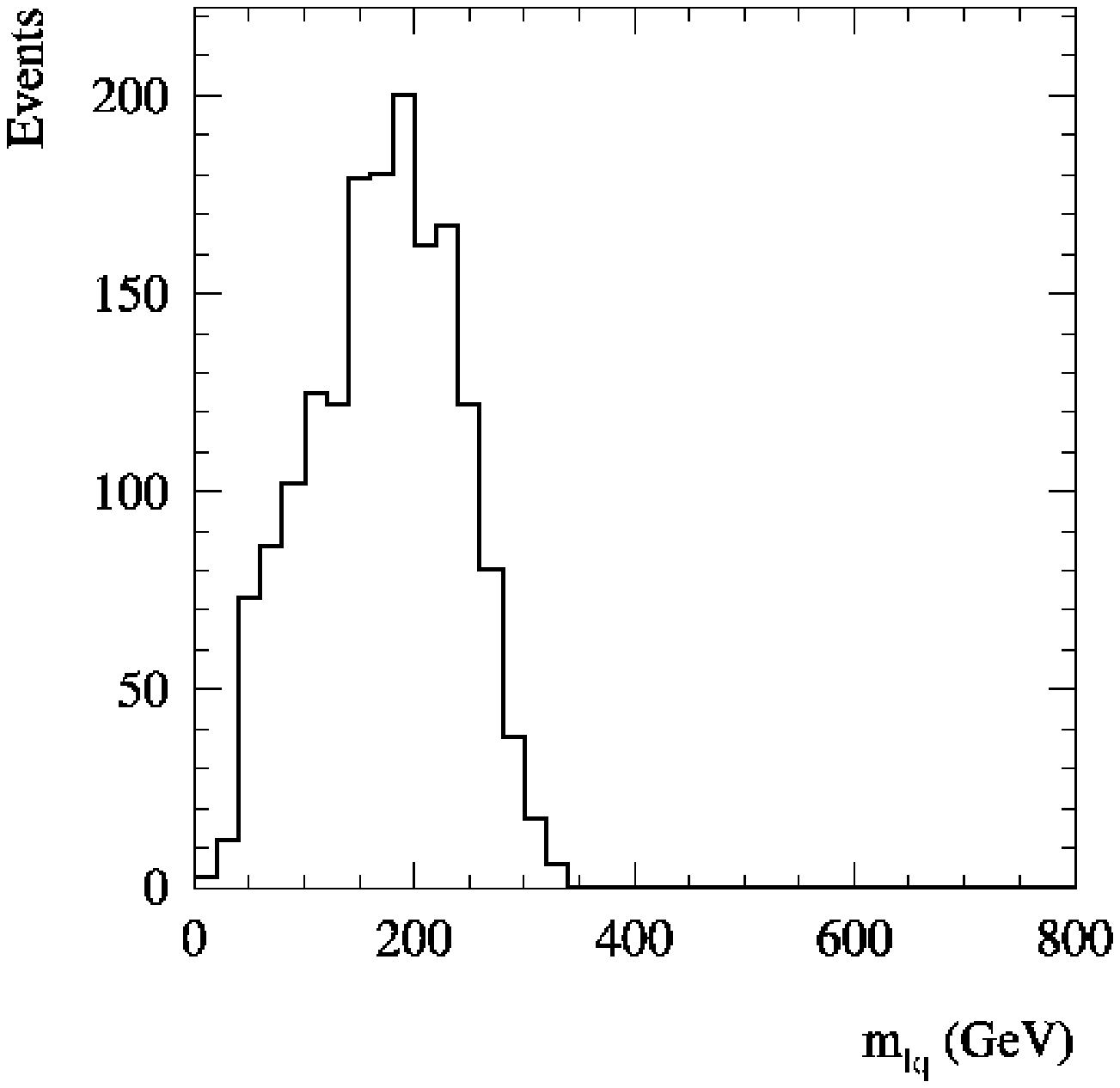,width=2.6in}{The flavour subtracted $m_{lq}^{high}$ distribution \label{mlqhigh} for the NUHM model defined in the text, constructed by taking the jet (from the two hardest jets in the event) that gives the lowest $m_{llq}$ invariant mass and forming the highest invariant mass that one can make with the two leptons in the event.}{The flavour subtracted $m_{lq}^{low}$ distribution \label{mlqlow} for the NUHM model defined in the text, constructed by taking the jet (from the two hardest jets in the event) that gives the lowest $m_{llq}$ invariant mass and forming the highest invariant mass that one can make with the two leptons in the event.}

\DOUBLEFIGURE[t]
{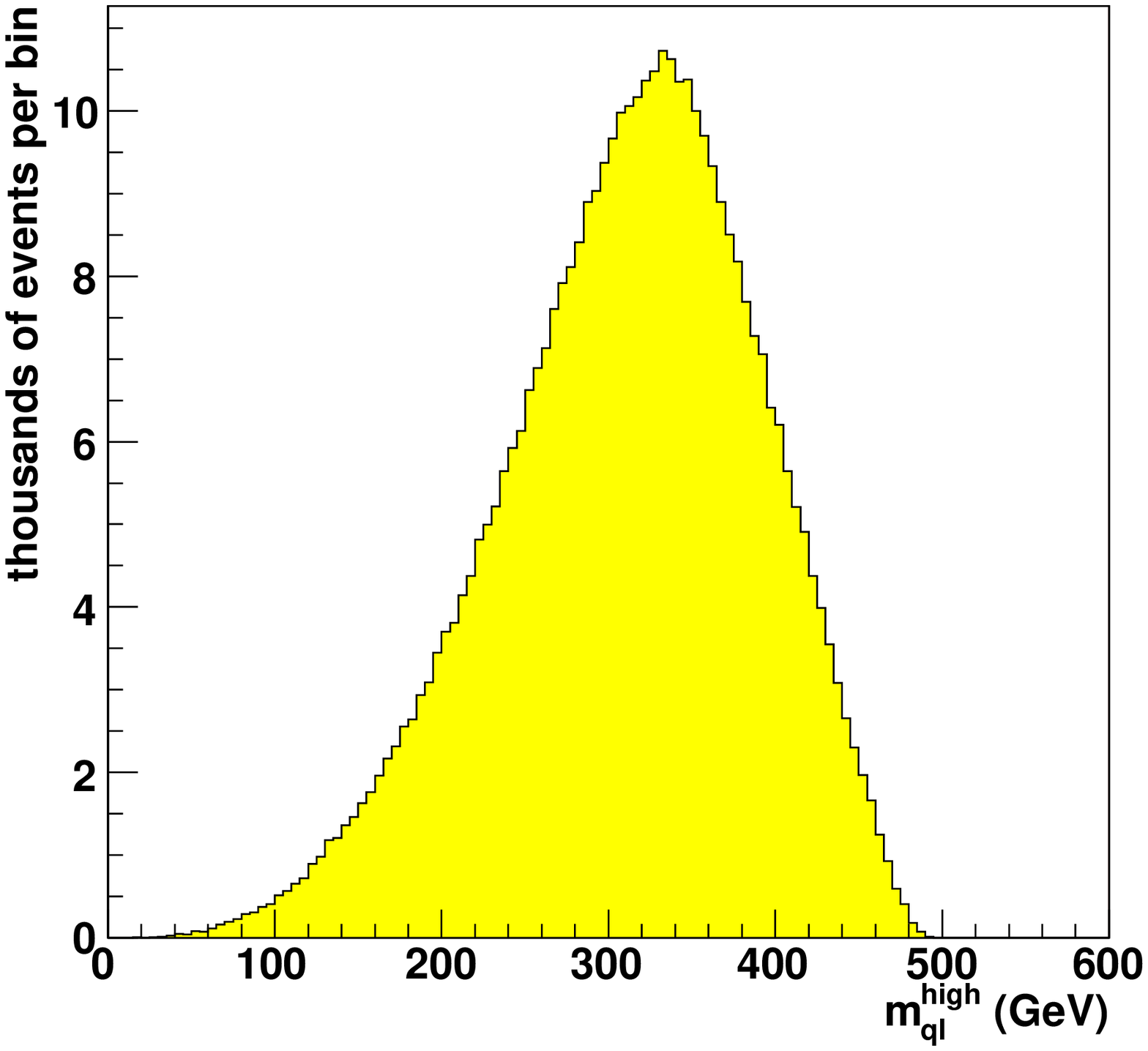,width=2.6in}
{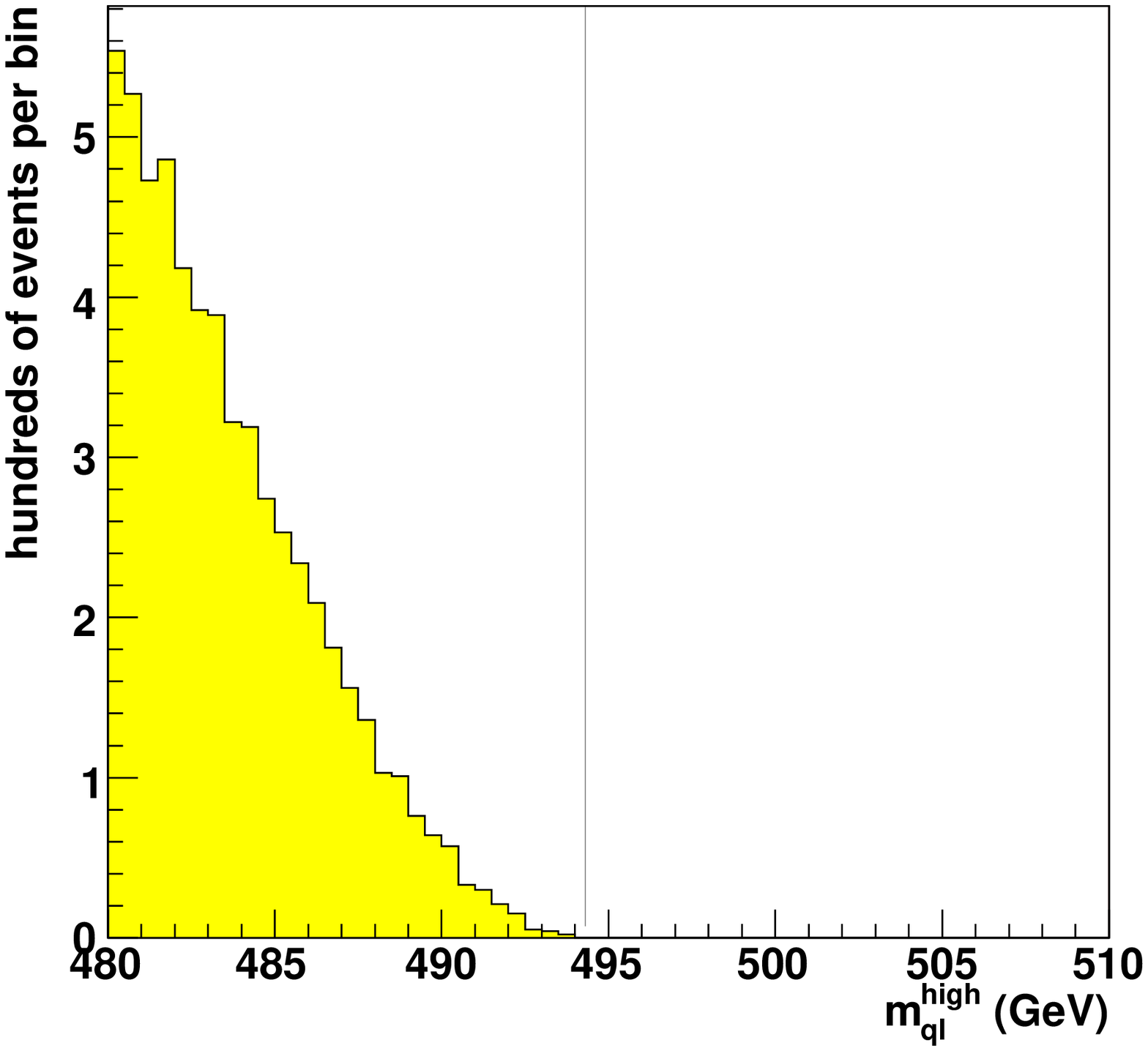,width=2.6in}
{A toy Monte Carlo calculation of the shape of an example $m_{lq}^{high}$ distribution
  \label{mlspace} (for particle masses of $m_\ntlinoOne=95$~GeV,
  $m_\ntlinoTwo=179$~GeV and $m_\squark=610$~GeV similar to those in the NUHM model
  defined in the text) using the approximation in which all particles
  are taken to be scalars, i.e.\ phase-space only.  The calculation
  assumes 100\% acceptance and does not model any detector
  effects. $m_{lq}^{high}$ values are in GeV.}{A
  \label{mlspacezoom} zoomed view of the $m_{lq}^{high}$ phase space
  distribution (left) in the vicinity of the end point
  $(m_{lq})^{max}_{high}$, the position of which is given by the
  vertical line at 494.24~GeV. $m_{lq}^{high}$ values are in GeV.}

\section{Discussion}\label{discussion}
Having collected a series of endpoint plots and derived their positions in terms of the sparticle masses involved in our decay chain, we ought in principle to be able to reconstruct the masses in the chain. This is not as simple as it first appears, however, for the following reasons:
\begin{enumerate}
\item
Our NUHM point is in the mass region where the $m_{lq}^{high}$ endpoint is in the same position as the $m_{llq}$ endpoint, and we also know that the $m_{lq}^{low}$ edge is related to these other two merely through multiplication by a constant factor. Discounting the poorly measured threshold, we thus only have two equations in three unknowns (we can use the $m_{ll}$ endpoint as our second equation), and we do not have enough information to constrain the masses.
\item 
The observation of endpoints does not reveal anything about the decay process that produces them, and the shapes of the lepton-jet invariant mass distributions in this paper are not dissimilar to those produced by chains of successive two body decays. Hence, we need to consider how we would in principle distinguish between the two cases to be certain that we are reconstructing the correct masses.
\end{enumerate}
Given that the purpose of this note is simply to present a derivation of the three body endpoint expressions and demonstrate their existence in a SUSY model, we will not implement solutions to either of these problems. However, the following discussion will suggest possible answers to both.
\subsection{Mass reconstruction}
In order to reconstruct the sparticle masses in our process, we will need to supply extra constraints. One possible method involves going one step higher in the decay chain, and searching for decays of the form:
\begin{equation}
\tilde{g}\rightarrow \tilde{q}q \rightarrow \tilde{\chi}_2^0qq \rightarrow \tilde{\chi}_1^0 qq ll
\end{equation}
This will give us more endpoints, and hence we will obtain direct constraints on the masses of the system. This is similar to an approach used for two body decays in \cite{Gjelsten:2005aw}, and would have the advantage of providing a measurement of the gluino mass. The problem here is that our NUHM model has a relatively light gluino, which is lighter than the squarks of the first two generations. Hence, the method will only be applicable to decay chains involving stop and sbottom squarks, and we see that whilst this approach may be helpful in some cases it is certainly not generally applicable to all regions of parameter space. 

Another approach is to use other observables to constrain sparticle masses. Our previous work in \cite{Lester:2005je} demonstrated that one can obtain a dramatic improvement in mass measurements by combining exclusive data (such as endpoint information) with inclusive observables (such as the cross-section of events passing a missing $p_T$ cut, to give one example). This analysis could be repeated here, and with enough inclusive observables one could obtain good measurements of the sparticle masses involved in our cascade process. This has the advantage of being generally applicable regardless of the mass spectrum, though it relies on specifying a particular SUSY model in which to perform the analysis.
\subsection{Decay chain ambiguity}
We have already remarked that it is not trivial to distinguish three body endpoints from two body endpoints, and hence the observation of endpoints alone is not enough to be able to reconstruct masses. Even assuming that we $do$ have a way to do this, we can in general exchange the neutralinos in our decay process for other neutralinos without changing the observed signature, and so one cannot assume that we know exactly which process is causing the observed edges. This latter point was previously considered by us in \cite{Lester:2005je} for the case of two body decays, and here we will concern ourselves with the former question of distinguishing between two and three body decays.

In the previous sections we have already seen one characteristic feature of three body decay chains; the ratio of the $m_{llq}$ and $m_{lq}^{low}$ endpoint positions is always $\sqrt{2}$. Provided one can obtain precise measurements of these quantities, we would have a clue that we were looking at three body processes, although this could easily be faked by two body decays that conspired to produce endpoints in similar positions.

For an extra clue, consider that although the shapes of the lepton-jet distributions in three body decays are not dissimilar to those encountered in two body processes, this is not true of the $m_{ll}$ distribution, and hence there is potentially some information contained in the shape of the dilepton distribution. In the two-body case, one obtains a triangular shape that is identical to the phase space distribution. In contrast, the three-body distribution in figure \ref{mll-plot} is heavily peaked toward the endpoint. Unfortunately, this is unlikely to be true over the whole of parameter space; the three body decay proceeds via an off-shell Z or slepton, and the $m_{ll}$ distribution is peaked toward the endpoint when the endpoint (which is the same as the mass difference between the $\tilde{\chi}_2^0$ and the $\tilde{\chi}_1^0$) approaches, for example, the Z mass, such that the shape will depend heavily on the SUSY point. Furthermore, a previous study investigated the effect on the $m_{ll}$ shape of incorporating the matrix element for the three body decay process in addition to the phase space, for different points in mSUGRA parameter space, and although for some points it dramatically altered the pure phase space result there were other points where the three body and phase space shapes were virtually indistinguishable \cite{Chouridou:2003aw}.

To summarise, we may be fortunate enough to find that nature presents a point at which we can distinguish three body decays from two body processes simply by looking at the endpoint shapes and positions, but this is certainly not guaranteed. For this reason, a better approach to the problem of mass reconstruction is to use the Markov Chain techniques presented in \cite{Lester:2005je}, where no assumption is made about the processes causing the observed endpoints. This allows us to select a region of the parameter space consistent with the data which can be used as a basis for further investigation.

\section{Conclusions}\label{conclusions}
We have derived expressions for the position of the kinematic endpoints arising in cascade decays featuring a two body decay followed by a three body decay, and have applied them to the lepton-jet invariant mass distributions given by a squark decay process. We have performed the first analysis of an NUHM model as it would appear within the ATLAS detector, and, using standard cuts, have observed endpoints that are consistent with the predicted positions. We thus conclude that the technique is a viable extension of the current method used in chains of two body decays. 

We have discussed the problem of mass reconstruction in models with a similar phenomenology, and have found that it is hampered by both a lack of constraint from the endpoint equations themselves, and the problem of distinguishing three body from two body decay processes. In the case of the NUHM benchmark point $\gamma$, one would be able to identify the decays as three body decays by using the shape of the dilepton invariant mass distribution. 
\section{Acknowledgements}
We owe thanks to the other members of the Cambridge Supersymmetry
Working Group, in particular Peter Richardson for information
regarding the {\tt HERWIG} treatment of three-body neutralino
decays. We acknowledge Borge Gjelsten, whom we understand has
independently calculated endpoints for three body decays, but has not
yet published them.  This work has been performed within the ATLAS
Collaboration.  We have made use of {\tt ATLFAST} which is the result
of collaboration-wide efforts within ATLAS.  We thank ATLAS
collaboration members for useful discussions, and in particular wish
to thank Paul de Jong and an anonymous ``ATLAS Scientific Note
referee'' for bringing important issues to our attention.

\bibliography{testbib}
\end{document}